\documentclass[twocolumn]{aastex62}
\usepackage{amsmath,amssymb,natbib}

\shorttitle{Spectral Characteristics of WINQSEs}
\shortauthors{Mondal et al.}

\begin{document}

\title{Characterizing the Spectral Structure of Weak Impulsive Narrowband Quiet Sun Emissions}

\correspondingauthor{Surajit Mondal}
\email{surajit.mondal@njit.edu}

\author[0000-0002-2325-5298]{Surajit Mondal}
\affiliation{Center for Solar-Terrestrial Research, New Jersey Institute of Technology, 323 M L King Jr Boulevard, Newark, NJ 07102-1982, USA}

\author[0000-0002-4768-9058]{Divya Oberoi}
\affiliation{National Centre for Radio Astrophysics, Tata Institute of Fundamental Research, S.P. Pune University Campus, Pune 411007, India}

\author[0000-0002-1741-6286]{Ayan Biswas}
\affiliation{National Centre for Radio Astrophysics, Tata Institute of Fundamental Research, S.P. Pune University Campus, Pune 411007, India}
\affiliation{Department of Physics, Engineering Physics \& Astronomy, Queen’s University, Kingston, Ontario K7L 3N6, Canada}
\affiliation{Department of Physics, Royal Military College of Canada, Kingston, Ontario K7K 7B4, Canada}

\author[0000-0001-8801-9635]{Devojyoti Kansabanik}
\affiliation{National Centre for Radio Astrophysics, Tata Institute of Fundamental Research, S.P. Pune University Campus, Pune 411007, India}



\begin{abstract}
Weak Impulsive Narrowband Quiet Sun Emissions (WINQSEs) are a newly discovered class of radio emission from the solar corona. These emissions are characterized by their extremely impulsive, narrowband and ubiquitous nature.
We have systematically been working on their detailed characterization, including their strengths, morphologies, temporal characteristics, energies, etc.
This work is the next step in this series and focuses on the spectral nature of WINQSEs. 
Given that their strength is only a few percent of the background solar emission, 
we have adopted an extremely conservative approach to reliably identify WINQSES.
Only a handful of WINQSEs meet all of our stringent criteria.
Their flux densities lie in the 20 $-$ 50 Jy range and they have compact morphologies.
For the first time, we estimate their bandwidths and find them to be less than 700 kHz, consistent with expectations based on earlier observations.
Interestingly, we also find similarities between the spectral nature of WINQSEs and the solar radio spikes. This is consistent with our hypothesis that the WINQSEs are the weaker cousins of the type-III radio bursts and are likely to be the low-frequency radio counterparts of the nanoflares, originally hypothesized as a possible explanation for coronal heating.
\end{abstract}


\section{Introduction}
\label{intro}

Weak Impulsive Narrowband Quiet Sun Emissions (WINQSEs) are a recently discovered class of solar radio transients \citep[][henceforth referred to as M20]{mondal2020}. WINQSEs were also discovered in a quieter time using independent analysis techniques \citep[][]{sharma2022, bawaji2022, mondal2023}. While the literature about solar radio transients in general is a fairly mature field, not much is known about WINQSEs. Both M20 and \citet{mondal2023} (henceforth referred to as M23) demonstrated that WINQSEs are highly impulsive in nature, with most of them lasting for a fraction of a second. M20 and \citet{mondal2021} postulated that WINQSEs are the radio counterparts of the ``nanoflares" hypothesized to explain coronal heating \citep{parker1988}. They proposed that small-scale magnetic reconnection events happening in the corona induce magnetohydrodynamic waves which then propagate up along magnetic field lines and often induce even smaller magnetic reconnections higher up in the corona. These magnetic reconnections also produce nonthermal electron beams which then emit plasma emission in the radio band due to their interaction with the thermal plasma. The key difference between these weak flares and the standard type-III solar radio bursts is that the type-III radio bursts are much brighter and often outshine the quiet solar disc. The nonthermal electron beams responsible for producing the type-III radio bursts { often have speeds $\sim 0.3-0.4c$,} generally travel for large distances, and give rise to emissions often spanning hundreds of MHz.
However, since the beams responsible for producing WINQSEs are expected to be very weak energetically, they cannot traverse large distances. Consequently, WINQSEs are expected to have small bandwidths. It is reasonable to expect from the long tail seen in their distribution towards greater flux densities (M20 and M23), that there will be a handful of WINQSEs which are produced by electron beams energetic enough to traverse larger distances. These WINQSEs will then give rise to radio signatures resembling weaker type-IIIs, { or type-Is for that matter,} in the dynamic spectra. The first such feature{, resembling a weak type-III,} has been reported by \citet{sharma2022}.
They found an emission feature consistent with an electron beam speed of $\sim0.1c$ and a peak brightness temperature of only $\sim60 kK$.

By comparing the radio and the EUV images visually, provided in \citet{sharma2022}, it seems that the source of the electron beam is a coronal bright point, which are known to be miniature active regions, and powered by small-scale magnetic reconnections. That the electron beams produced from these regions { are} stronger than that produced in the quiet sun is consistent with expectations. 

A similar picture was also proposed to explain the radio emission from a microflare \citep{mohan2019b}. A rough estimate of the strengths of the associated electron beams can be obtained by comparing the strengths of the radio emissions. 
If one assumes that the radiative efficiency of the electron beams does not vary widely over the range of strengths of electron beams considered here,
then the strength of the observed radio emission would be directly proportional to the energy content of the electron beams. 
\citet{mohan2019b} reported a brightness temperature of $\sim 10^8$ K of the radio emissions associated with the microflare. 
Overlapping in time, a group of type-III radio bursts was in progress at lower radio frequencies, for which \citet{mohan2019a} reported a brightness temperature of $10^9$ K. 
By comparison, the brightness temperature of WINQSEs is barely above $10^6$ K (as can be deduced from what was reported by M20). Hence it follows that the electron beams responsible for producing WINQSEs of strengths similar to that observed in M20 and M23 
are much weaker than those responsible for the emission observed by \citet{mohan2019b}. This, in turn, implies that the bandwidth of the WINQSEs must also be much smaller than the 7-8 MHz reported by \citet{mohan2019b}. 
However, the current upper limit on bandwidth, $\lesssim 12$MHz, reported by M20, is insufficient to test this hypothesis. \citet{sharma2022} have also not commented on the spectral nature of the general population of WINQSEs. 
The objective of this paper is to push the available data to its limits to quantify the spectral nature of WINQSEs and check if the expectation of their bandwidths being much smaller than the current upper limit is borne out.


The paper is organized as follows -- Section \ref{sec:observation} describes the observations and data analysis strategy, Section \ref{sec:results} presents the results and finally in Section \ref{sec:discussion}, we discuss the implications of this work and the path forward.

\section{Observation and data analysis}\label{sec:observation}

\subsection{Observation Details}
The radio data presented in this work were recorded with the Murchison Widefield Array \citep[MWA;][]{lonsdale2009, tingay2013, wayth2018} on 2020 June 20 between 04:16:00 $-$ 04:17:00 UTC under project code G0002. 
The MWA design allows one to distribute its 30.72 MHz of instantaneous observing bandwidth flexibly in the 80 $-$ 300 MHz observing band of the instrument.
This is achieved by first doing a coarse filtering of the Nyquist sampled data over 256 spectral channels, each with a bandwidth of 1.28 MHz. 
Of these, 24 coarse channels are transported to the correlator, housed in a central processing facility, where each of them is subjected to a further filtering operation yielding fine spectral channels of bandwidth 10 kHz.
As in the rest of the MWA literature, we refer to the 1.28 MHz channels obtained after the first stage of filtering as {\em coarse} channels and the 10 kHz channels obtained after the second stage of filtering as {\em fine} channels.
The implementation details of the channelization and the MWA correlator are provided in \citet{tingay2013}. Due to this frequency architecture, the MWA band suffers from high rms at the edges of coarse channels and the central fine channel of a coarse channel \citep[e.g.][]{offringa2015,beardsley2016,wice2016}. Data from these channels are not suitable for scientific investigation. This makes MWA incapable of continuous and uniform sampling in the frequency space.

\begin{figure}
    \centering
   \includegraphics[clip,trim={5cm 0.5cm 3cm 1cm},scale=0.44]{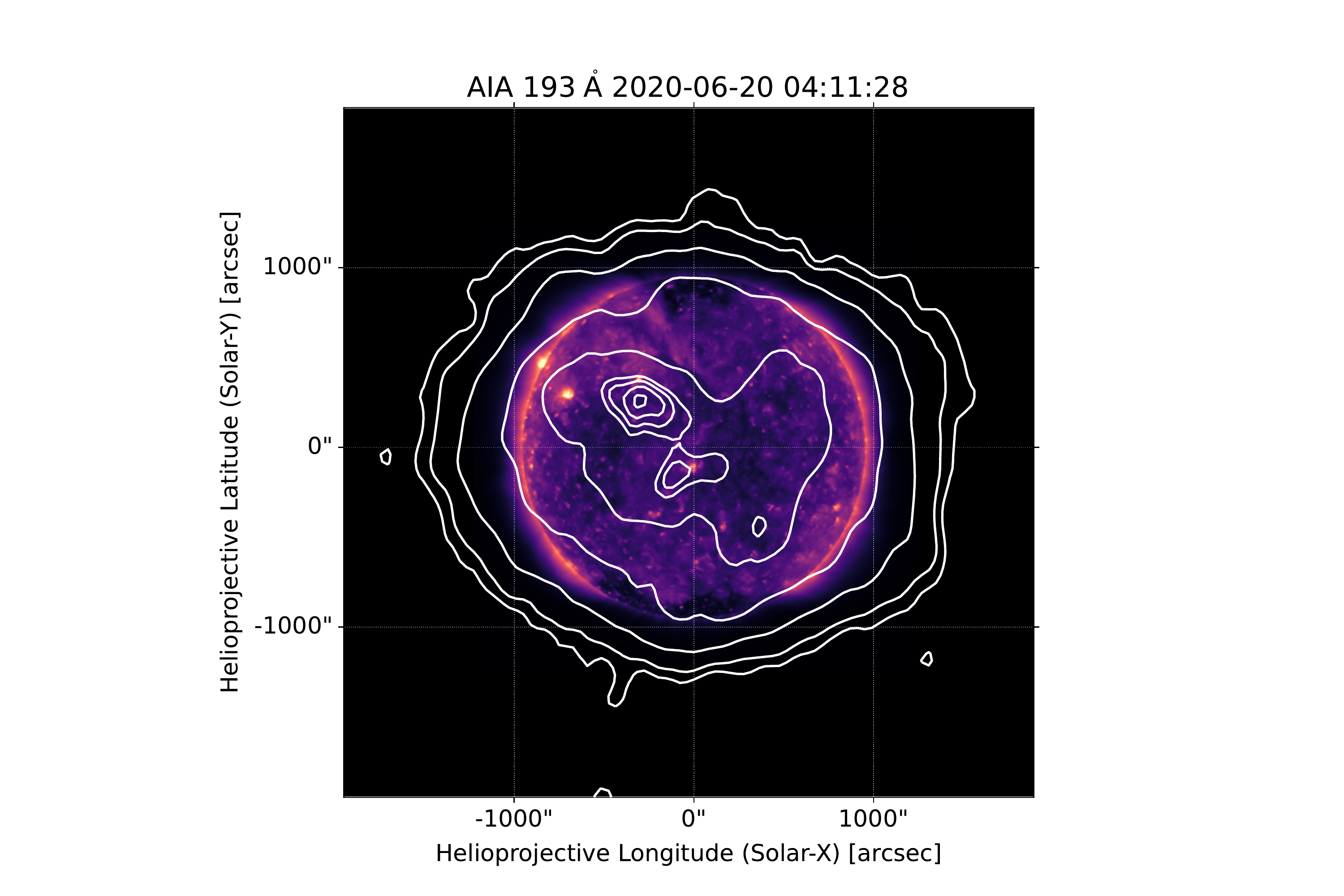}
    \caption{The contours show an example 133 MHz radio image at 04:11:24.5 UTC, overlaid on an AIA 193 \AA $\,$ image taken near the middle of the observations. The contours are at 0.04, 0.08, 0.2, 0.4, 0.6, 0.72, 0.74, 0.76, 0.78, 0.8, 0.9 times the peak in the radio image.}
    \label{fig:aia_images}
\end{figure}

The frequency of this observation spanned $\sim$120 $-$ 150 MHz. The data were taken with a time resolution of 0.5s. For ease of data handling, the data were averaged to 40 kHz resolution. For context, contours of an example radio image are overlaid on an 193 \AA$\,$ filter of the Atmospheric Imaging Assembly \citep[AIA;][]{lemen2012}, onboard the Solar Dynamics Observatory \citep[SDO,][]{Pesnell2012}, during the observing period are shown in Figure \ref{fig:aia_images}. The structure of the Sun is very similar at other radio frequencies. As is evident from these figures, no sunspot was present on the visible part of the solar disc. The sun was extremely quiet on this day. Additionally, no radio flare was reported during the observing window by the Learmonth Spectrograph, which operates in the range 25 $-$ 180 MHz.

\subsection{Calibration and Imaging}

Here we present an analysis of MWA observations in the frequency range between 121--145 MHz, covered by $\sim$19 coarse spectral channels, each of width 1.28 MHz.
Standard flagging was performed to remove RFI and instrumental artifacts. Calibration was done using the Automated Imaging Routine for Compact Arrays for Radio Sun \citep[AIRCARS;][]{mondal2019}. 

While AIRCARS has the capability to calibrate every time and frequency slice independently, this is very time consuming and often not required. Additionally, in absence of any strong compact source on the solar disc, data from the long baselines often have a very low signal-to-noise ratio (SNR) leading to poor calibration solutions for the outer antennas in the case of the MWA. 
Additionally, determining the calibration solutions for each 40 kHz time slice is highly time consuming, due to the sheer number of spectral channels involved. 
With these considerations in mind, we have chosen a 160 kHz chunk every 1.28 MHz and obtained the calibration solutions for it using data averaged over 9 s. The calibration solutions are applied to the 0.5 s time resolution and 40 kHz frequency resolution data using linear interpolation in time and assuming that the calibration solutions do not change over a coarse channel. The validity of this assumption will be discussed later. The final images are made at 160 kHz frequency resolution and 0.5 s time resolution. In total we have made $\sim$ 18000 images for this project. The brightest feature on the Sun was detected typically at a signal-to-noise ratio of $\sim 60$. Next we smoothed all images using a circular beam of angular resolution of $280$ arcsec. We chose to work with a circular beam to remove any effects of the orientation of the point-spread-function from the final images. Images at this resolution are used for all further analysis, unless mentioned otherwise. For more details about the imaging and calibration steps, we refer the readers to M23. 

\subsection{Flux-density Calibration} \label{sec:flux}
Since the goal of this investigation is to study the spectrum of WINQSEs, it is imperative that images at all frequencies must be brought to the same flux density scale. 
Fortuitously, these solar observations included Crab in the MWA field-of-view and its spectrum was used for flux density calibration.
The flux densities of the Crab at multiple frequencies were obtained from the NASA Extragalactic Database and a powerlaw was fit to the spectrum. This best fit powerlaw was used as the reference for the flux density of Crab at different frequencies. 
We extracted the integrated flux density of Crab from the images produced using the task \textit{imfit} implemented in the Common Astronomy Software Applications \citep[CASA,][]{mcmullin2007,CASA2022}. A median over time was taken to smooth out the effect of temporal variability. The flux density scaling factor at frequency $f$, denoted by $\alpha_{scale}(f)$, and the absolute flux density of Sun at frequency $f$, denoted by $S^{absolute}_{sun}(f)$, are given by,
\begin{equation}
    \alpha_{scale}(f)= \frac{F^{absolute}_{crab}(f)}{F^{image}_{crab}(f)}\times P_{crab}(f)
\end{equation}
and
\begin{equation}
      S^{absolute}_{sun}(f)= F^{image}_{sun}(f)\times \alpha_{scale}(f)\times \frac{1}{P_{sun}(f)},
\end{equation}
where $F^{absolute}_{crab}(f)$ and $F^{image}_{crab}(f)$ denote the absolute flux density of Crab at frequency $f$ obtained from the powerlaw fit and the median flux density of Crab at frequency $f$ obtained from the image respectively. Similarly  $F^{image}_{sun}(f)$ is the flux density of the Sun at frequency $f$ obtained from the image. $P_{crab}(f)$ and $P_{sun}(f)$ are the value of the primary beam gain at frequency $f$ towards the direction of the Crab and the Sun respectively. 
We use the Full Embedded Element MWA primary beam model provided by \citet{Sokowlski2017}.
It has been shown by \citet{kansabanik2022a} that a similar procedure can also be used to obtain the bandshape of an instrument.

\subsection{Obtaining Spectra of WINQSEs} \label{sec:obtain_spectrum}
Conventionally, in radio interferometry instrumental bandpass calibration solutions are derived with high accuracy for spectral analysis work.
A continuum map is then produced after applying these calibration solutions. If the spectral line is reasonably strong, the continuum map is produced by excluding the frequencies where the spectral line is expected, under the assumption that the frequency structure of the sky is smooth, once the spectral lines have been removed. The continuum map is used to generate the model visibilities. These model visibilities are then subtracted from the calibrated visibilities to produce residual visibilities which are then used to produce a spectral image cube. The spectrum at the location of interest in the image can then be extracted from the cube. A low-order polynomial is then fitted to the spectrum excluding the spectral line and subtracted. The spectrum obtained after this step is regarded as the final spectrum. 
\begin{figure*}
    \centering
    \includegraphics[trim={1.5cm 0 2.7cm 0},clip,scale=0.55]{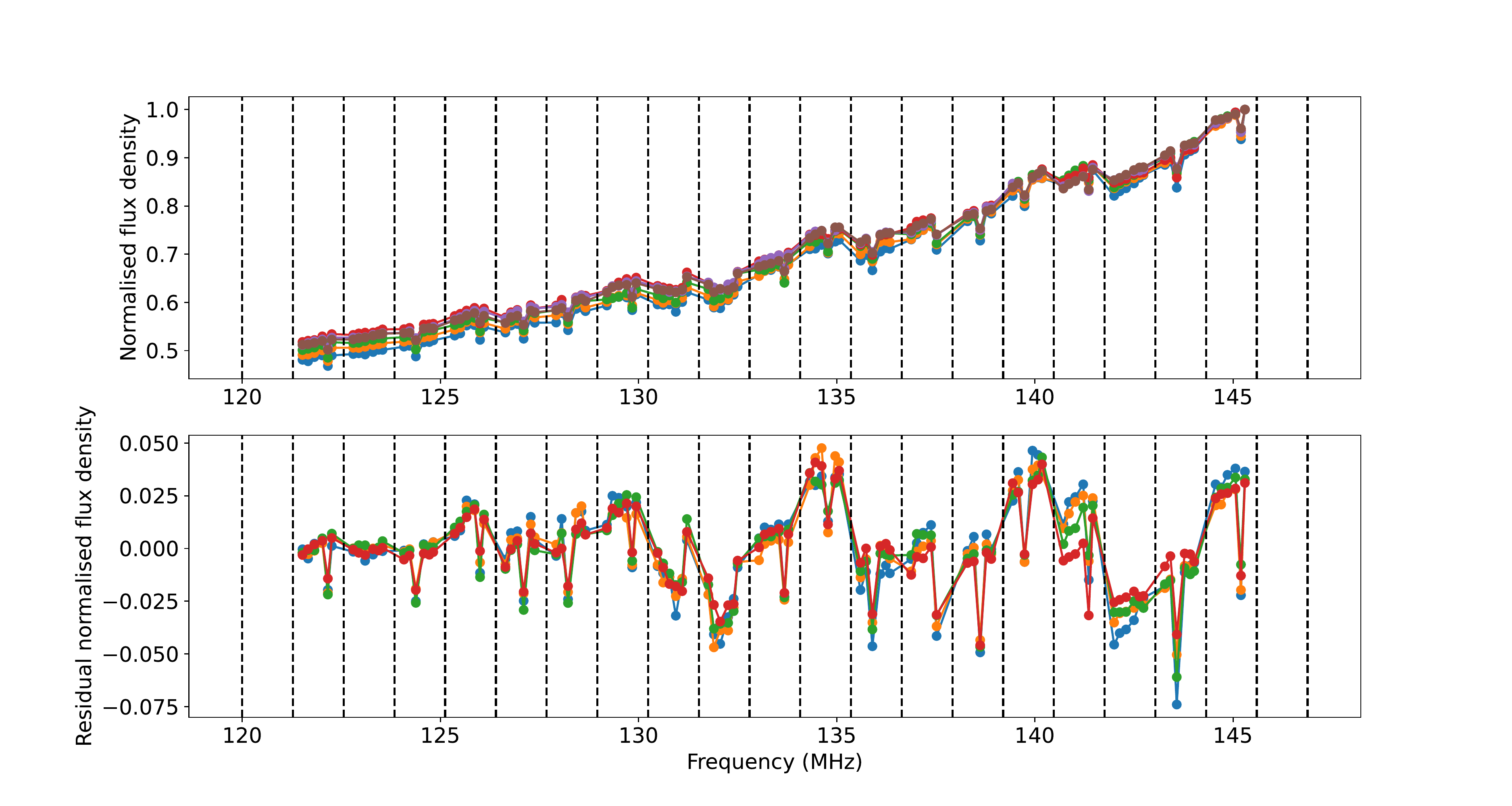}
    \caption{Upper panel: Normalized median spectra for a few pixels on the solar disc. Bottom panel: Residual spectra after subtracting the fitted polynomial corresponding to the normalized median spectra shown in the left panel.
     The vertical dashed lines mark the boundaries of the coarse channels used in this observation.
    }
    \label{fig:median_map}
\end{figure*}

The times, frequencies, and locations of WINQSEs are unpredictable (M20 and M23). 
They have high occurrence rates, take place all over the band, and are ubiquitous on the quiet Sun.
It is hence not possible to {\em a-priori} identify times, frequencies, or solar locations which are assured to be free of WINQSEs.
The conventional radio interferometric spectral analysis approach, which is based on knowing the spectral and spatial location of the spectral feature, is hence not suitable in the present context.
To make progress on this, we make some simplifying assumptions.
Given that their spatio-temporal fractional occupancy is less than 15\% (M23), 
we can safely assume that the median solar map over time at every frequency can be treated as a true representation of the ``quiet" (WINQSEs free) Sun at that frequency. Hence subtracting the median map at every frequency is similar to doing a continuum subtraction. The upper panel of Figure \ref{fig:median_map} shows the normalized median spectrum from a few different pixels on the Sun, which rises by a factor of two over this short bandwidth.
It is evident from this figure that all fine spectral channels within a single coarse channel have very similar median flux densities, validating the assumption that the bandshape does not vary significantly within a coarse channel. Let us denote the normalised median spectrum as $S^{normalized}$. While qualitatively $S^{normalized}$ is similar at different pixels, there are small differences across different pixels. To show these small differences, we fit a quadratic polynomial to $S^{normalized}$ from each pixel. Let us denote the resultant quantity as $S_{fitted}^{normalized}$. The difference between $S_{fitted}^{normalized}$ and $S^{normalized}$ is shown in the bottom panel of Figure \ref{fig:median_map}. For each line shown in the upper panel of Figure \ref{fig:median_map}, there is a corresponding line in the bottom panel. 
It is evident that the residuals are small (generally not larger than $\pm4$\%) and their variation with frequency is qualitatively and quantitatively similar for all the pixels.
We find that one channel out of the six 160 kHz channels 
plotted here for every coarse channel sometimes 
usually show significantly different residual from the other five channels. 
As shown later, the strength of the features of interest here is much larger than that of these artifacts.
Hence, their presence does not pose much challenge in the present context.
A detailed understanding of the variations seen in the residuals is yet to be developed. 
Though unrelated,
spectral features of similar strength 
were reported by \citet{wice2016} in bandshapes of individual tiles and were attributed to reflections in the cable bringing the signal from the tiles to the receivers in the field. 

For every timeslice, we subtracted the median image from the image of that time and frequency slice and obtained the residual image. Then we convolved the image with a circle of diameter $280$ arcsec, which is equal to the angular resolution.
The pixel value which is in units of $Jy/beam$, where beam refers to the restoring beam, was then divided by the beam area. 
This was done to ensure that the numerical value of a pixel becomes equal to the flux density from an area equal to the angular resolution and centred on that pixel. 
This dataset (henceforth referred as the residual time-frequency hypercube) was used for further analysis.
We refer to a slice of the residual time-frequency hypercube along the temporal axis as residual cube.
We also computed the rms of these maps to get an estimate of the noise in the maps. 

The spectrum for every pixel on the Sun was extracted from the residual cube for every time slice and the following steps were performed to determine if it met the criteria to be regarded as a significant detection of a WINQSE.
Given their low flux densities, the individual detection of WINQSEs can only be expected to be at comparatively low SNRs. 
However, by requiring multiple independent criteria to be met simultaneously, one can bolster the confidence in the reliability of the detection and characterization of WINQSE and their spectrum.
    \begin{figure*}
    \centering
    \includegraphics[trim={0cm 0cm 1.7cm 0cm},clip,scale=0.63]{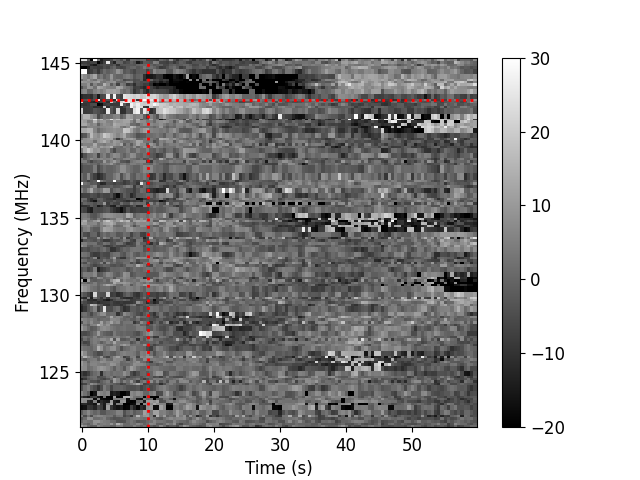}\includegraphics[trim={0cm 0cm 0cm 0cm},clip,scale=0.63]{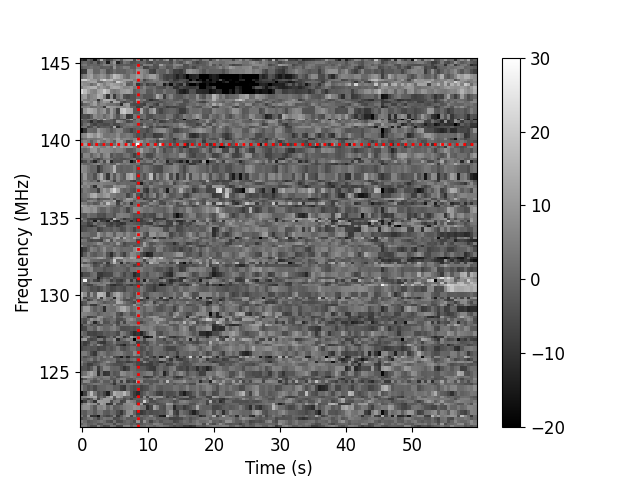}
    \caption{
    Median subtracted dynamic spectra for some example pixels showing strong emissions. The features of interest lie at the intersection of the red dashed lines in both panels. 
    The left panel shows a strong feature lying in a band of artefacts, and hence is not regarded to be a WINQSE.
    The right panel shows a strong feature which lies far away from the few artefacts present in the dynamic spectrum and is regarded as a bonafide WINQSE.
    }
    \label{fig:dynamic_spectrum_bad}
\end{figure*}

\begin{enumerate}
    \item Smooth the spectrum with a running mean filter of lengths 1, 3, 5, till 15 in steps of 2. At each step we calculate the SNR, which is defined as the ratio of the peak in the smoothed spectrum to the rms in the smoothed spectrum. If the SNR drops at any step we stop the iteration in filter lengths and the previous filter length is noted down as the bandwidth. We denote this bandwidth by $\Delta \nu$ and the location of the peak frequency as $\nu_0$.
    
    \item Fit a straight line to the spectrum, excluding the part between [$\nu_0-\Delta \nu/2-1$, $\nu_0+\Delta \nu/2+1$ ], to ensure that the peak {does} not influence the fitting in any way.
    The best fit straight line was subtracted from the entire spectrum to remove any residual slope in the spectrum.
    
    \item We classify the spectrum as ``good" if it satisfies all of the following conditions: 
    
    \begin{enumerate}
         \item The peak of the spectrum lies 10 or more spectral channels away from the edge of the frequency band analyzed here. This ensures that there is sufficient data on either side of the peak to reliably estimate the spectral baseline.

        \item The peak of the smoothed baseline subtracted spectrum (henceforth referred to as smoothed spectrum) is greater than 4 times the rms of the smoothed spectrum. \label{condition1}
        
        \item The peak of the smoothed spectrum is greater than 4 times the rms in the residual map corresponding to $\nu_0$  for the timeslice under consideration. \label{condition2}
        
        \item The peak of the smoothed spectrum is greater than 1.5 times the absolute value of the minimum in the spectrum. \label{condition3}
        
    \end{enumerate}
    
    \item We then group the spectra thus obtained using a friends-of-friends algorithm subject to the condition that two spectra are ``friends" of each other if they are separated by at most 280 arcsec in both the spatial dimensions, by at most 1 spectral channels of width 160 kHz in the frequency dimension and coincident in time. 
    The motivation behind this comes from the fact that compact emission features like the WINQSEs are expected to be clustered at the scale of the restoring beam and might be correlated across the frequency axis as well. Noise peaks on the other hand are independent across frequency and it is unlikely that multiple noise peaks lying within a restoring beam will satisfy the stringent criterion laid out above.
   In practise, a typical number of friends for spectra are about 10, and they all tend to lie in the same spectral slice.

    \item The spectrum with the highest SNR within a group of ``friends" is chosen to be its representative.
    
    
    \item We verify by visual inspection that there are no artefacts in the dynamic spectrum of strengths comparable to the peak of the representative spectrum and also lying close to the peak in the dynamic spectrum of median subtracted flux density for that particular pixel. If this condition is not true, we discard those spectra. One example of such an artefacts is shown in the left panel of Figure \ref{fig:dynamic_spectrum_bad}. A  WINQSE satisfying the criteria just mentioned is shown in the right panel of the same figure. The feature under consideration is located at the intersection of the two dotted red lines in both the panels. It is evident that the dynamic spectrum in the left panel shows many artefacts, most notably horizontal bands which map on to coarse channels in the MWA band. The feature of interest in the dynamic spectrum lies in one such band and hence we discard this feature. The right panel also shows some artefacts, in particular a strong negative band, but located far away from the feature of interest. Hence we regard this feature as a reliable detection of a WINQSE. 
    
    \item We generate a moment zero map corresponding to the remaining spectra. The procedure for generating the moment zero map is as follows:
    \begin{enumerate}
        \item Let $\nu_0^i$, $\Delta \nu^i$ and $t^i$ be the peak frequency, bandwidth and time of occurrence respectively, for a representative spectrum.
        \item Consider the set of spectral slices from the residual image cube corresponding to time $t^i$. 
        \item Individually fit a straight line to the spectra for every pixel while excluding the frequencies between $[\nu_0^i-\Delta \nu^i/2-1$, $\nu_0^i+\Delta \nu^i/2+1]$, and subtract the best fit straight line from the entire spectrum for the corresponding pixel.
        \item Average the baseline subtracted spectrum over frequency between $[\nu_0^i-\Delta \nu^i/2$, $\nu_0^i+\Delta \nu^i/2]$.
    \end{enumerate}

    \item In addition to the previous three criteria (Criteria \ref{condition1}, \ref{condition2}, \ref{condition3}), we regard the representative spectrum as ``good", if the following two conditions are also met:
    \begin{enumerate}
        \item The pixel value in the moment zero map at the spatial location of the representative spectrum is greater than the absolute value of the deepest negative in the moment zero map. \label{condition4}
        \item The moment zero map does not contain any peak outside the solar region greater than the pixel value in the moment map at the spatial location of the representative spectrum. The solar region is defined to be a square of size $3.0 R_\odot \times 3.0 R_\odot$ centred on the Sun. \label{condition5}
        \end{enumerate}  
    
\end{enumerate}

We believe that at the end of this rather stringent filtering process, we are left with only reliable detections of WINQSE spectra. Hereafter, we refer to the spectra meeting these criteria as ``good" spectra. 
While it is very likely that the stringent criterion we employ will also reject some true WINQSEs, it is however the assurance that the ones which make it through are the ones which have been detected with high significance, which is more important here.


\section{Results}\label{sec:results}

\begin{figure*}
\includegraphics[trim={2cm 0 0 0},scale=0.6]{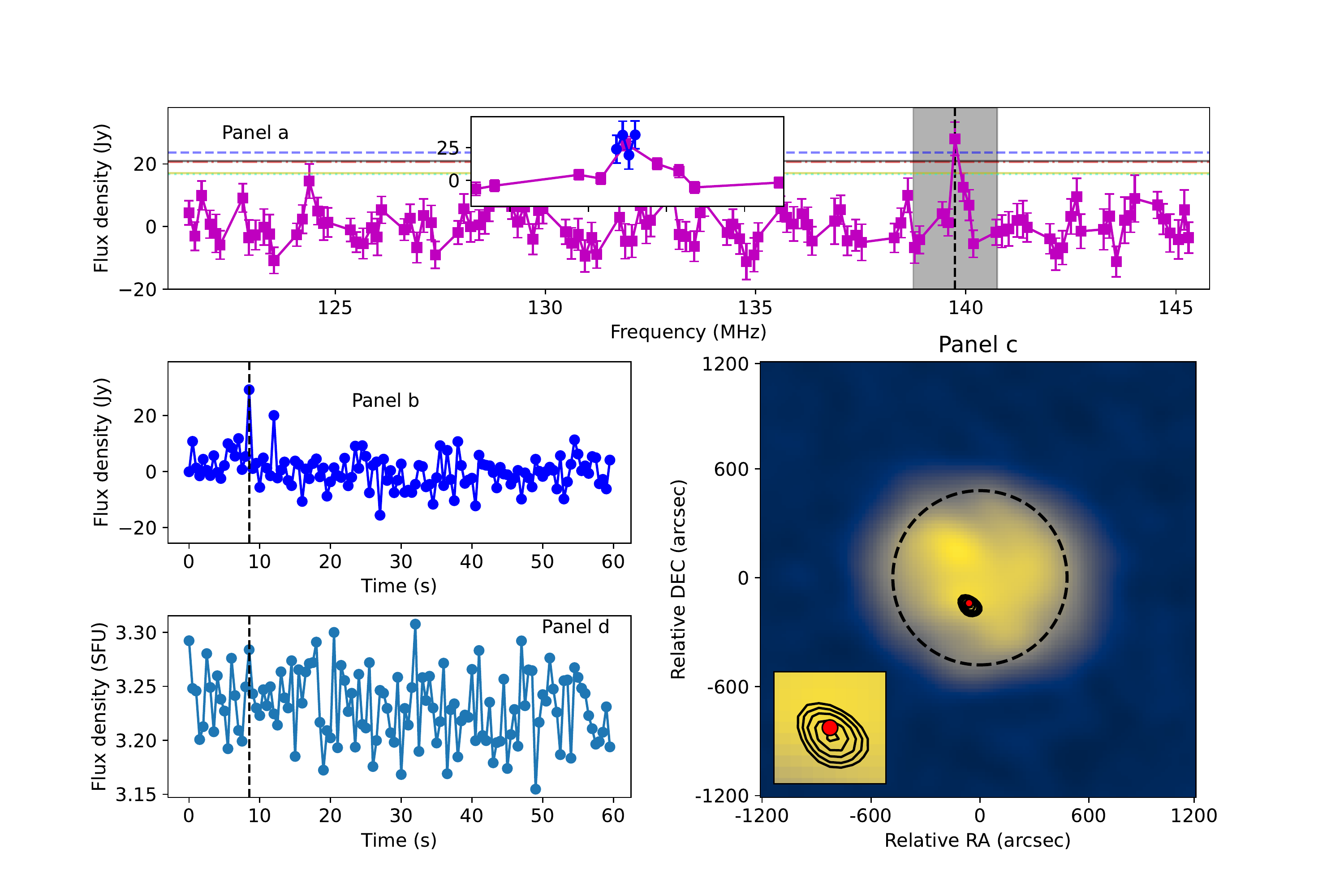}
\caption{{\bf Panel a:} Spectra at 160 kHz frequency resolution are shown in magenta. The blue points are at 40 kHz frequency resolution. 
The vertical black dashed line show the location of the spectrum peak. The horizontal lines show different thresholds used to determine if the spectrum is ``good". The blue dashed, red dot-dashed, cyan dotted, yellow solid and black solid lines represent thresholds corresponding to conditions \ref{condition1}, \ref{condition2}, \ref{condition3}, \ref{condition4} and \ref{condition5}, respectively, which are mentioned in Section \ref{sec:obtain_spectrum}. 
A zoomed in view is shown in the inset. The extent of the inset along the frequency axis is shown with gray shaded region.  {\bf Panel b:} The flux density timeseries extracted from the location of the WINQSE at the peak frequency. The dashed black line shows the time when the WINQSE was detected. {\bf Panel c:} Contours on the moment map for this WINQSE are overlaid on the map corresponding to the peak frequency of this spectrum. The contour levels are chosen at -0.99, -0.95, -0.9, -0.85, -0.8, 0.8, 0.85, 0.9, 0.95, 0.99 times the peak in the moment map. However, no negative contour is present. The zoomed version of this map at the location of the WINQSE is shown in the inset. {\bf Panel d:} The time series of integrated flux density of the whole Sun. The online material shows properties of all the ``good" spectrum in this same format.
}
\label{fig:spectra}
\end{figure*}

We find five spectra meeting the stringent criteria laid out in the previous section.
Their peak flux densities lie between 20 $-$ 50 Jy. 
We do not detect any spectra with bandwidth greater than 700 kHz. 
Multiple spectra were unresolved at 160 kHz spectral resolution, primarily because of gaps in our spectral sampling due to instrumental artefacts. 
To investigate if these spectra can be resolved with 40 kHz frequency resolution, we made 40 kHz maps for the time slices where WINQSEs were detected. We assumed that the median spectrum (described in Section \ref{sec:obtain_spectrum}) and also the flux scaling factor does not change across these 160 kHz. 
The spectra at 40 kHz resolution were extracted in a manner analogous to that for spectra at 160 kHz resolution. 
We found all of the WINQSEs to be resolved at the 40 kHz frequency resolution.
Details of one such WINQSE spectrum are shown in the panel a of Figure \ref{fig:spectra} and the rest are available in the supplementary material. 
The spectrum and the flux density times series at the spatial location of WINQSE are shown in panels a and b of Figure \ref{fig:spectra}, respectively, clearly showing both its narrowband and impulsive nature. 
The moment map corresponding to the spectrum in panel a is shown 
in panel c of the same figure. 
It clearly shows that there is neither any noise present at the level of the WINQSE, nor is there any negative contour at a comparable level. 
Panel d shows the variation of integrated flux density of the whole Sun with time at the peak frequency. 
The $\Delta F/F$ for the spectrum shown in Figure \ref{fig:spectra} is $\sim 0.7$, where $\Delta F/F$ is the ratio of the flux density of the WINQSE to the median solar flux density. Note that this is significantly larger than the $\pm4$\% residuals seen in Figure \ref{fig:median_map}.

We note the following limitations of the current data and the analysis. 
These data often tend to show presence of artefacts which map on to coarse channels width (1.28 MHz).
Many of the high SNR features detected early on in the analysis needed to be discarded because they were found to lie in or close to a part of the dynamic spectrum which showed significant artefacts (Figure \ref{fig:dynamic_spectrum_bad}).
Additional limitations arise due to the gaps and non-uniformity in spectral sampling. These come about because the edges of the coarse channels and the central fine spectral channel of each coarse channel need to be discarded as they suffer from instrumental systematic.
While our analysis yielded no WINQSEs with bandwidths larger than 700 kHz, it is useful to remain mindful that these data are not well suited for estimating bandwidths of weak emissions at spectral scales close to the widths of the coarse channels.

\section{Discussion}\label{sec:discussion}

We have presented a detailed analysis to carefully identify WINQSEs in the MWA solar data and estimate their flux densities, spectral characteristics and morphology.
Reliable characterization of weak, short-lived and narrowband features is challenging and care is needed for separating artefacts and noise peaks from true features of interest.
Towards the end, we devised a set of stringent selection criterion which required a given feature to simultaneously satisfy a large number of independent constraints which a true feature can be expected to satisfy.
We find five WINQSEs which meet these stringent criterion. 
Their flux densities lie in the range of 20 $-$ 50 Jy. 
They are all narrowband with maximum widths $700$ kHz and many with widths $\sim160$ kHz. All of these spectral features were resolved at a frequency resolution of 40 kHz.
This is consistent with the hypothesis that the electron beams responsible for WINQSEs are much weaker than those responsible for the events studied by \citet{mohan2019b}, and are hence also much smaller in their spectral span.

Interestingly we find that the bandwidths and duration of the WINQSEs studied here are very similar to that of solar radio spikes \citep[][]{clarkson2021,clarkson2023}. This observation is also consistent with our hypothesis that WINQSEs { arise from electron beams much weaker than those giving rise to the typical type-III or type-I radio bursts.
}.  From the moment map, it is evident that the WINQSEs are compact in nature, which is also consistent with the hypothesis put forth in \citet{mondal2021} and observationally confirmed by \citet{sharma2022} and \citet{bawaji2022} using independent analysis techniques. 

{
We note that the hypothesis of WINQSEs being weak type-IIIs or type-Is only draws upon the fact that all these emission arise from coherent emission mechanisms, which at our observing frequencies is generally plasma emission.
In this sense they bear similarities to other emissions arising from similar mechanisms, like striae and radio spikes etc.
The observed details of these features will naturally depend upon the speed of the electron beams involved and the orientations of the magnetic fields along which they travel with respect to the local plasma density gradient.
}


M21 stressed the need to search for the counterparts of WINQSEs in other wavebands to understand the relationship between them and also to verify if these are indeed the radio counterparts of the long hypothesised nanoflares.  The extremely narrowband and impulsive nature of the WINQSEs suggest that for finding WINQSE corresponding to events observed in other energy bands, search will need to be done at high spectral and temporal resolution over very large bandwidths. A more practically feasible approach will be to first detect WINQSEs in the radio bands and then try to identify their counterparts in other wavebands.

As mentioned earlier, the key limitations of this work comes from the spectral systematics 
in the MWA data. The MWA is currently going an ambitious upgrade, referred to as MWA Phase-III. 
The MWA Phase-III will use new receivers with improved digital signal processing which have been carefully designed to reduce the spectral systematics. It will also have 256 antenna tiles, twice the present number.
The additional antennas will quadruple the number of instantaneously available baselines, significantly boosting the sensitivity and the imaging performance of the array.
These improvements will allow us to overcome the most constraining limitations of the present effort.
It is also worth mentioning here that search for similar weak transients from the quiet Sun is being enthusiastically pursued across multiple wavebands including higher radio frequencies \citep[e.g.][]{nindos2020,nindos2021,panesar2021,mandal2021,mondal2023b,alissandrakis2023}. Together these observations probe the solar atmosphere ranging all the way from the chromosphere to the corona. 
Combining the information across these observations will yield a more complete understanding of the low level dynamics ubiquitous on the Sun and hopefully bring us closer to a resolution of the coronal heating problem.

\facilities{Murchison Widefield Array (MWA) \citep{lonsdale2009,tingay2013}, Atmospheric Imaging Assembly \citep[AIA;][]{lemen2012} of the Solar Dynamics Observatory \citep[SDO;][]{Pesnell2012}}

\software{Astropy \citep{astropy:2013,astropy:2018}, Matplotlib \citep{Hunter:2007}, Numpy \citep{harris2020array}, SciPy \citep{2020SciPy-NMeth}, CASA \citep{mcmullin2007,CASA2022}, AIRCARS \citep{mondal2019}.}

\begin{acknowledgements}
This scientific work makes use of the Murchison Radio-astronomy Observatory (MRO), operated by the Commonwealth Scientific and Industrial Research Organisation (CSIRO).
We acknowledge the Wajarri Yamatji people as the traditional owners of the Observatory site. 
Support for the operation of the MWA is provided by the Australian Government's National Collaborative Research Infrastructure Strategy (NCRIS), under a contract to Curtin University administered by Astronomy Australia Limited. We acknowledge the Pawsey Supercomputing Centre, which is supported by the Western Australian and Australian Governments.
S.M. acknowledges partial support by USA NSF grant AGS-1654382 to the New Jersey Institute of Technology.
D.O., A.B. and D.K. acknowledge support of the Department of Atomic Energy, Government of India, under the project no. 12-R\&D-TFR-5.02-0700.
This research has also made use of NASA's Astrophysics Data System (ADS). 
\end{acknowledgements}

\bibliography{bibliography}   

\end{document}